\documentstyle[12pt,aas2pp4]{article}
\begin{document}
\newcommand{\kms}{\,km~s$^{-1}$}      
\newcommand{\subsun}{\mbox{$_{\odot}$}}
\newcommand{\Msun}{M\subsun}
\newcommand{\Rsun}{R\subsun}

\def\ltsima{$\; \buildrel < \over \sim \;$}
\def\simlt{\lower.5ex\hbox{\ltsima}}
\def\gtsima{$\; \buildrel > \over \sim \;$}
\def\simgt{\lower.5ex\hbox{\gtsima}}
\def\arcs{$''~$}
\def\arcm{$'~$}

\title{Blue Horizontal Branch Stars in M92\altaffilmark{1}}

\author{\sc J.G. Cohen\altaffilmark{2} and J. K. McCarthy\altaffilmark{2}}

\altaffiltext{1} {Based in large part
on observations obtained at the W.M.Keck Observatory,
which is jointly operated by the California Institute of Technology
and the University of California.}

\altaffiltext{2} {Palomar Observatory, Mail Stop 105-24,
California Institute of Technology}

\begin{abstract}
We have analyzed high dispersion and high precision spectra of 
5 blue horizontal
branch stars in the globular cluster M92 to 
establish that the projected rotational
velocity for these stars ranges from 15 to 40 \kms.  This is
larger than that expected based on the rotation of their main
sequence progenitors, the spin down of rotation with age, and
the conservation of angular momentum.  Possible explanations
include a rapidly rotating stellar core.

An abundance analysis of these spectra of these blue HB stars in M92 
yields the same results as have been obtained from the giants in this cluster.
There is a hint of a trend of higher abundance as the projected surface
rotational velocity increases, which could be chance and requires confirmation.

\bigskip
\bigskip
\bigskip
\bigskip
\bigskip
\bigskip

\centerline {{\it Subject headings:} stars: rotation --- stars: chemical
composition --- globular clusters: M92}
\bigskip
\bigskip
\bigskip
\bigskip
\bigskip
\bigskip
\centerline{In Press in {\sl The Astronomical Journal.}}

\end{abstract}

\section{INTRODUCTION}

There is a large suite of complicated puzzles involving
the properties of the horizontal branch in globular clusters.
One of these involves the distribution of stars along the horizontal
branch, and how that distribution varies from cluster to cluster, i.e.
the second parameter problem. (See, for example, Fusi Pecci et al 1992, 1993.)
We believe that the mass loss
varies from star to star along the horizontal branch in a given globular
cluster, but do not understand what parameters control the mass loss
near the tip of the giant branch and at the helium core flash.
Another puzzle involves the 
possible impact of the mixing of the lighter elements, C, N, and O,
and probably Na, Mg, and Al as well, produced by nuclear burning
in the interior of these evolved stars near the tip of the
red giant branch to their surfaces, as reviewed by Kraft (1994).
We do not know how, or even if, that is related in some way
to the morphology of the horizontal branch.  There have also
been tantalizing hints of unexpectedly large surface rotational
velocities among the
horizontal branch stars (Peterson 1983, 1985 a,b),
but the data available
have until recently been very limited. 

If conservation of angular momentum prevails,
high surface rotation among the red giants in a globular cluster
would not be expected 
due to their large radii, nor is it
observed.  There is essentially no information on
the surface rotation of lower main sequence stars in globulars.  High
internal rotation would be impossible to detect among lower main sequence
stars or among red giants, but might be revealed after mass loss
at the helium flash when the stars reach the horizontal branch.  It is
this phenomenon which we may have detected, although other interpretations
are of course possible.

As
discussed by  
Sweigart \& Mengel (1979) and by Sweigart (1996),
high rotation may influence the process and outcome of
mixing, halt diffusion of elements out of the atmosphere, 
affect the population of the horizontal branch, etc. 

The advent of the Keck 10-m Telescope, and its high resolution
echelle spectrograph (Vogt et al 1994), prompted us to re-examine
some of these issues.

\section{OBSERVATIONS}

Five blue horizontal branch stars in M92 that were in the outer parts
of the cluster (so they would be well isolated spatially) were chosen 
for study.  Charts and U, B, V photometry can be found in 
Sandage \& Walker (1966) and Sandage (1969).
These objects were observed using the HIRES spectrograph
on the Keck-1 Telescope
in 1994 October.  The slit used was 1.1 arc-sec wide and 7 arc-sec
long, which gives a spectral resolution of 34,000.
The detector is a 2048 x 2048 pixel Tektronix CCD with a readout noise
of 5 electrons rms.  The projected slit corresponds to 4 pixels in
the dispersion direction.  The exposures
were binned by a factor of 2
in the direction along the slit prior to readout.
Three 1200 sec exposures were obtained for each star; they were
analyzed, sky subtracted and summed to form the final spectra.
Even at such high dispersion,
sky subtraction is particularly vital in the region of the NaD lines. 
The data were reduced using Figaro.

The spectra cover the wavelength range from 4304 to 6630$\AA$~ (orders
82 to 54).  With the current HIRES CCD detector,
there is full spectral coverage at the blue end ($\lambda < 5000\AA$), but
gaps of about 30\% of the apparent order length develop at the red end.

The radial velocity stability of HIRES is excellent as it is permanently
mounted in a temperature stabilized housing on one of the Nasmyth
platforms of the Keck-1 Telescope.  A Th-Ar lamp was used for wavelength
calibration.

The precision of these data is very high.  The signal-to-noise ratio,
measured by looking at the statistics in continuum segments 5\AA~ long
near the center of each order
that appear to be line free, is between 40 and 70 per pixel 
(80 to 140 per resolution
element).  This is an
underestimate because of the probable presence of undetected weak
absorption features within each segment.

\section{ROTATIONAL VELOCITIES}

In all earlier work, mostly by Ruth Peterson and her collaborators
(Peterson 1983, 1985 a,b) 
where the spectral resolution 
and the signal-to-noise ratio were lower, 
the rotational broadening had to be inferred from
the width of the peak of the cross correlation between the spectrum
of a rotating star and that of one  believed to have little or no rotation.
Peterson, Rood \& Crocker (1995) succeeded in resolving the lines of
the infrared OI
triplet in blue HB stars in several globular clusters, but because
the OI triplet lines are on the saturated part of the curve of growth,
the determination of rotation is coupled to the abundance of oxygen, which
varies widely from star to star in globular clusters as described
by, for example, Sneden et al 1991.
Hence the solution for $v_{rot}$ becomes quite complex.
However, in our case the profiles of much weaker absorption lines could
be detected, and it was apparent from an initial inspection
of the raw data at the telescope as well as of the reduced
HIRES spectra
that the rotational velocities were large and differed significantly
from star to star.  Figure 1 illustrates the comparison over a 24\AA~ region
of the spectrum of the
most rapidly rotating M92 horizontal branch star in our sample
(M92 XII--1) to that of
one of the slowly rotating stars.  The equivalent width of the weakest
line marked in each panel, the 5167\AA~ MgI line, is 39 m\AA~ for M92 X--22
and 86 m\AA~ for M92 XII--1.  The spectra are normalized to the continuum,
whose value in DN (1 DN ${\approx}~2 e^-$) is given in the panels of the figure.
The mean FWHM of isolated
lines in the Th-Ar arc is $8.6 \pm0.3$ \kms. 

%
%

In our spectra, in part because of the very low metallicity of 
M92, the absorption
lines are weak and in most cases well isolated.  The stellar 
absorption lines which were used
to determine the rotation ranged in equivalent width from 20 to 120 m\AA.
This means that these lines are largely on the linear part of the
curve of growth, and that their intrinsic profiles are identical, scaled
of course to varying total absorption.  Damping parameters and abundances
of the line absorber are irrelevant to the shape of the normalized line profile.

Two methods of determining the rotational velocity were used.  The
first relied on measuring the width of the best fitting Gaussian
profiles for a number of isolated, unblended absorption features.
Between 5 and 15 lines were used for each star.
These were compared to similar measurements of the profiles 
of the arc emission lines, broadened
by varying amounts of rotational broadening. 
The rotational broadening profile adopted is
that of Uns\"old (1955). A value for the limb darkening
coefficient of $\beta$ = 2.7 was used, based on Allen (1973).
The results are insensitive to a change in $\beta$ by a factor of two.
This method did not
work well for the most rapidly rotating stars, where a Gaussian
does not provide an adequate fit to the observed line profiles.

%
%

The second method
was to choose a set of four reasonably strong 
stellar absorption lines in each star, shift and scale
their profiles appropriately to form a composite line profile for
each M92 HB star, $Y_{star}$.  The appropriate scaling was done
in the wavelength domain as well.
The result of this process was
then compared with the profile of a single Th-Ar arc emission
line, broadened
by varying amounts of rotation, $Y(arc)_{broad}$.
A statistical measure of the agreement of the
two profiles was computed, although the final rotational velocity
assigned using this method was chosen by eye after examining plots
with $v_{rot}(sin(i))$ near that giving the minimum error.
An example of this for the M92 blue HB star in our sample with the
largest projected $v_{rot}$ is shown in Figure~2.  The lines included,
with their equivalent widths in M92 XII--1,
are 4383.6 (Fe I, 86 m$\AA$), 4404.8 (Fe I, 55 m$\AA$), 
4468.5 (Ti II, 108 m$\AA$), and 4572.0$\AA$(Ti II, 77 m$\AA$).
A four point running mean of the shifted, scaled, and overlaid profiles
is displayed.

The results are listed in Table~1.  The first column gives the
star name (from Sandage \& Walker 1966), the second gives the
rotational velocity inferred from the Gaussian fits, the third column
gives the rotational velocity inferred from the composite line
profile fitting followed by its error.
The errors in the first case are the 1$\sigma$
rms errors for the group of lines, while in the second case the errors
are based on 
$ \Sigma {\lbrack}Y(arc)_{broad} - Y_{star}{\rbrack}^2 $
evaluated at 0.01\AA~ intervals over the
region with absorption greater than 3\% of the continuum, calculated for
each value of $v_{rot}$.
The agreement between the two independent 
determinations of $v_{rot}(sin(i))$ is gratifying.

\begin{deluxetable}{lccccc}
\tablewidth{0pc}
\scriptsize
\tablecaption{Projected Rotational Velocities for Blue 
Horizontal Branch Stars in M92}
\tablehead{
\colhead{Star} & \colhead{$v_{rot}(sin(i))$} & 
\colhead{$\sigma$} & \colhead{$v_{rot}(sin(i))$} &
\colhead{$\sigma$} & \colhead{$V_r$} \nl  
\colhead{ } & \colhead{Gaussian} &  \colhead{ } & \colhead{Profile Fit} &
\colhead{  }  &  \colhead{(\kms)} \nl
\colhead{ } & \colhead{(\kms)} & \colhead{(\kms)} & \colhead{(\kms)} & 
\colhead{(\kms)}}
\startdata
IV$-$17  &   14 & 2      &   15  & 4  &   $-$127.6  \nl
IV$-$27  &   22 & 3      &   27  & 4  &   $-$115.7  \nl
X$-$22   &   13 & 2      &   15  & 3  &   $-$121.6  \nl
XII$-$1  &   37 & 4      &   43  & 5  &   $-$127.6  \nl
XII$-$9  &   22 & 3      &   29  & 3  &   $-$127.7  \nl
\enddata
\end{deluxetable}

There is no correlation between projected rotational velocity and
$T_{eff}$.

We would like to know if the distribution of $v_{rot}(sin(i))$ is consistent
with a constant value of $v_{rot}$ and a random distribution of the angle
of inclination of the axis of rotation with respect to the line of sight.
A $\chi^2$ test applied to these data indicates a probabilty of 
$\approx$10\% that the sample of blue HB stars in M92 is derived from
such a population with 
a uniform rotation of about 45 \kms.
A larger sample could refine (or refute) this statement.

The final column of Table~1 gives the heliocentric radial velocities.
As expected,
all the stars are members of M92 with $<v_r> = -123.7$ \kms~
and $\sigma$ = 5.1
\kms.  This velocity dispersion is in excellent agreement with that
determined by Lupton et al (1985) from a much larger sample of stars in M92.

\section{DISCUSSION OF THE ROTATIONAL VELOCITIES}

The main sequence progenitors of the blue horizontal branch stars
in M92 have a mass of about 0.85 \Msun~ and are late F to early
G dwarfs.  The rotation of main sequence stars has been a subject
of interest since the work of Kraft (1967).  For many nearby
stars, periods are now available
which yield the rotation velocity itself rather than the projected
velocity of rotation.   These have come from extensive monitoring 
of the
variability of the chromospheric emission in the core of the H and K
lines (Baliunas \& Vaughan 1985).
Benz et al (1984) showed that in clusters as young as the Hyades,
$<v_{rot}>$ is less than 12 \kms~ for 10 stars with spectral types
ranging from F8V to G5V.  $v_{rot}$ as large as 50 \kms~ is not reached in
the Hyades until the spectral type is F5V, too early to be the progenitors
of the blue HB stars in M92.  We assume that metallicity differences
do not significantly perturb this picture.

In addition, it is clear from studying
open clusters of varying ages younger than the Hyades
that there is a spin-down of rotation with
time for these stars, where the rotational decay due to magnetic
braking is predicted to be of the form $t^{-0.5}$ (Skumanich 1972,
Endal \& Sofia 1981).  Whether and how far this extends beyond the
age of the Hyades is not known, but
extrapolating blindly from 7 x 10$^8$ years (the determination
of the age of the Hyades by Mazzei \& Pigatto 1988) to ages characteristic
of globular
clusters, one expects a further spin-down of the rotational
velocity of a factor of $\approx$4.

Based on models of horizontal branch evolution,
the mean mass of the M92 blue HB stars is about 0.70 \Msun, with a
scatter of perhaps 0.03 \Msun.  (Crocker \& Rood 1988, Lee et al 1990).
We assume solid body rotation and ignore any further spin-down for the
main sequence stars
beyond that shown in the Hyades.  We
assume
that the angular momentum/gm in the ${\approx}0.15${\Msun} 
lost during stellar evolution between the main sequence and the HB
is the same as that of the initial star.  The radii
of these blue HB stars are $\approx$3 \Rsun, and we find that
the expected rotational
velocity for blue HB stars should be under 10 \kms.
 
Our projected rotational velocities, as well as the latest work by 
Peterson et al (1995),
show that velocities exceeding those derived under the assumptions
above are common among the blue HB stars in globular clusters.
Among the many possibilities that could explain these results is that
main sequence stars have a more rapidly rotating core, and as mass is
lost, a higher surface rotation is seen.  The evolution of
differentially rotating stars is discussed by Pinsonnealt et al (1991).
Peterson et al (1995) gives a detailed discussion of the myriads
of other possible explanations.

Several recent studies at high
dispersion of field RR Lyrae stars (Clementini et al 1995,
Lambert et al 1996, 
Peterson et al 1996, Fernley \& Barnes 1996) 
demonstrate that high rotation is not seen
in any of these objects, and the sample of field RR Lyrae stars
observed at high dispersion is now
reasonably large.  The only time that line broadening can be detected
is near $\phi \approx$ 0.85, and this presumably arises from the effect
of shocks.  These variables are HB stars only slightly cooler
than the coolest of the blue HB stars in M92 that we examine here.
We speculate this is related to the fact that in one case we are looking
at field stars with a relatively young mean age, and in the other
we are looking at much older objects, although it is also possible that
globular cluster stars have more angular momentum ab initio than
do field halo stars.  This difference is quite puzzling.

\section{ATMOSPHERIC PARAMETERS}

In preparation for an analysis of the spectral features we need to
determine the atmospheric parameters for these stars.  Photoelectric 
$U$, $B$, $V$ photometry already exists (Sandage 1970), and we supplement that
with photometry at $K$ (2.2$\mu$) using the new infrared camera
(Murphy et al 1995) at the 60-inch telescope on Palomar Mountain.  These 
measurements were kindly made and reduced by Mike Pahre of Caltech.
An aperture 12 arc-sec in diameter was used.  The typical 
photometric uncertainty is
0.05 mag.
Assuming the metallicity is [Fe/H]=$-$2.35 dex (Cohen 1979),
the observed $B-V$ and $V-K$ colors, corrected for E($B-V$) = 0.02 mag
(Sandage 1969), were used together with the colors predicted
from Kurucz's (1993) grid of LTE model atmospheres.

This procedure yields the temperatures
given in Table~2.  The $T_{eff}$ values from $V-K$ are
given double weight in determining the $T_{eff}$ used in the analysis
because they are much less sensitive to small observational errors
in the photometry.  The observed $K$ magnitudes are listed in the
second column. 

The surface
gravities in Table~2 were calculated once $T_{eff}$ was determined, given that the
mass and luminosity of the stars were known.  (The distance modulus 
adopted was 7.85 kpc from Cohen 1992 and $E(B-V) = 0.02$ mag was used.)  

In addition, the profile
of $H_{\gamma}$ was available to confirm that the choice of
$T_{eff}$ and surface gravity
were reasonable.  While $H_{\beta}$ is within the wavelength range,
the constraints of defining
its continuum in this echelle format made it less reliable to use,
and $H_{\gamma}$, formed deeper in the atmosphere, should be the
better diagnostic.  The procedure for defining the continuum for the
broad Balmer lines
in this echelle format is described in detail in McCarthy \& Nemec (1997).

\begin{deluxetable}{lcccccccc}
\tablewidth{0pc}
\scriptsize
\tablecaption{Atmospheric Parameters for M92 Blue HB Stars}
\tablehead{
\colhead{Star} & \colhead{$K$} & \colhead{$T_{eff}(B-V)$} & 
\colhead{${\delta}T_{eff}(B-V)$\tablenotemark{a}} &
\colhead{$T_{eff}(V-K)$} & 
\colhead{${\delta}T_{eff}(V-K)$\tablenotemark{b}} &
\colhead{Adopted $T_{eff}$} &
\colhead{$T_{eff}(H_{\gamma})$} & 
\colhead{Log(g)} \nl
\colhead{ }    &  \colhead{(mag)} & \colhead{($^{\circ}$K)} & 
\colhead{($^{\circ}$K)} & \colhead{($^{\circ}$K)}    &
\colhead{($^{\circ}$K)} & \colhead{($^{\circ}$K)} & \colhead{($^{\circ}$K)} &
\colhead{(dex)} }
\startdata
IV--17 & 15.44 & 8750 & (+750,--200) & 10000 & (+1000,--750) & 9375 & $>8250$ & 
3.6 \nl
IV--27 & 14.57 & 7500 & (+300,--200) & 7625 & (+250,--250) & 7550 & 
7500 $\pm$100 & 3.1 \nl
X--22 & 14.46 & 7500 & (+300,--200) & 7450 & (+250,--250) & 7450 & 
7500 $\pm$100 & 3.1 \nl
XII--1 & 14.31 & 7400 & (+200,--200) & 7250 & (+250,--250) & 7325 & 
7250 $\pm$100 & 3.0 \nl
XII--9 & 14.36 & 7600 & (+200,--200) & 7375 & (+250,--250) & 7500 & 
7450 $\pm$100 & 3.1 \nl
%
\tablenotetext{a}{Uncertainty in $T_{eff}$ from $\delta(B-V) = {\pm}0.04$ mag.
{\hglue 1.0truein}
        $^{\rm b}$Uncertainty in $T_{eff}$ from $\delta(V-K) = {\pm}0.10$ mag.}
\enddata
\end{deluxetable}

In the temperature regime around $T_{eff} \sim 7500$K, the $H_{\gamma}$
profile is much more sensitive to small variations in $T_{eff}$ than 
to small variations in log(g).  
We define $W(H_{\gamma})$ as the
equivalent width of the Balmer line measured over the regime
from 1\AA~ from the line center (to avoid the core itself)
to 16\AA~ from the line center, beyond which the profile is very difficult
to determine due to the echelle format of the observations.  
In this
regime,
for an increase in $T_{eff}$ of 250K, $W(H_{\gamma})$ calculated
from the Kurucz grid of theoretical Balmer line profiles increases by
12\%, while it only increases by 3\% for $\Delta(log(g))$ = 0.5 dex.

Since the luminosity and mass of these blue HB stars in M92
are rather tightly constrained,
a change in $T_{eff}$ of 20\%, which would easily be detectable,
is required to change the surface gravity
by a factor of two.  We will see that in deduced parameters such
as the abundances, the temperature uncertainties
will dominate.
Table~2 lists in the penultimate column $T_{eff}$ determined
from $H_{\gamma}$ for the four cooler blue HB stars.  The
very good agreement
further validates the choice of atmospheric parameters given
in Table~2.
(The hottest star has much broader Balmer line profiles and the
continuum determination thus becomes more difficult.)

The distinction in color and in Balmer line profiles 
for a fixed $T_{eff}$ and log(g)
between the 
[Fe/H] = $-$2.0 dex and
$-$2.5 dex grid of models is not significant.

\section{ABUNDANCES ON THE HB OF M92}

\subsection{Motivation and History}

Abundance variations
in C, N and O have been known to occur within globular clusters
for many years and have always been ascribed to mixing.
Carbon et al (1982), Pilachowski (1988),
and Sneden et al (1991) have studied various aspects of this problem
for large samples of giants in M92.  It is now clear 
that  Na, Mg, and Al  variations also occur
among red giants within individual globular clusters (see, for example,
Pilachowski et al 1996, who discuss Na and Mg abundances in 130
giants in M13)
and that they too are the
result of mixing of nuclear-processed material from the stellar interior.
Until Denisenkov \& Denisenkova's seminal paper (1990), no one
understood how to produce Na and Al in such 
low mass and relatively unevolved stars.
(See also Langer, Hoffman \& Sneden 1993.)  The interaction
between internal rotation and mixing is a critical one (Sweigart 
\& Mengel 1979, Sweigart 1996), as is that between rotation and
diffusion in the photosphere.

Detailed abundance analyses of M92 giants for
the heavier elements, which should be unmodified from the original
formation of the globular cluster for stars not evolved significantly
beyond the He flash, have been performed by many groups,
beginning with Helfer, Wallerstein \& Greenstein (1959),
then Cohen (1979), Peterson, Kurucz \& Carney (1990),
Sneden et al (1991) for a few selected elements, and Armosky et al
(1994) for the heavy s-process elements.

We analyze the abundances of several elements in the blue HB stars in 
M92 for which we have HIRES spectra.  We do this not to contribute
to the abundance determination for M92, which we regard as well
established by the many previous investigations utilizing high
dispersion spectra and modern model atmosphere analysis techniques
applied to the red giants.  Instead we intend to demonstrate,
in preparation for moving to much more metal-rich globular clusters,
that reliable abundances can be obtained from these HB stars,
and that our normalizations and procedures are correct.
Furthermore we need to establish that in contrast to the situation
with the hotter blue HB stars such as was found by
Glaspey et al (1989), there is no evidence for
gravitational settling among these cooler blue HB stars.
Lambert et al (1992) have also explored these issues, recent models
for meridional circulation and diffusion are given by Charbonneau \& Michaud
(1991) among others, and Vauclair \& Vauclair (1982) review the
taxonomy of hot stars with peculiar abundances that are believed
to have arisen as a result of element segregation in stellar photospheres.

\subsection{Our Abundances for the Blue HB Stars in M92}

We have adopted the table of transition probabilities of Luck (1992)
updated to include the modifications for FeI suggested by
Lambert, Heath, Lemke \& Drake (1996) and those for FeII
from Bi\'emont et al (1991).  The atmospheric parameters adopted
are those of Table~2.  The Kurucz (1993) grid of model atmospheres was
used together with the MOOG code of Sneden (1975).  Using the
solar model supplied by Kurucz (1993) and the solar
equivalent widths of Moore, Minnaert \& Houtgast (1966), 
we can reproduce the solar abundances satisfactorily.

We do not use any non-LTE corrections.  Clementini et al (1995),
Lambert et al (1996) and Fernley \& Barnes (1996)
discuss this issue for field
RR Lyrae stars whose atmospheric parameters are fairly
close to those of our M92 blue HB stars. The corrections for
non-LTE ionization equilibrium are fairly small
($< 0.2$) dex, but may become larger with decreasing metallicity
as the photoionizing UV flux increases.  In any case, our final
results (Table~3) give no evidence that a substantial correction is necessary.
No corrections for hyperfine structure were applied to any lines either.
A micro-turbulence velocity of 2.0 \kms~ was adopted.

One of the 5 blue HB stars in M92 is significantly hotter than the
other four stars, and its lines are much weaker.  Since M92 is
a very metal poor globular cluster, this makes all the lines in that star
extremely weak.  We therefore
only consider the four cooler stars.
Table~3 gives the number of lines used, the mean abundance
for each element and the $\sigma$ about the mean
for the abundance of each element
derived from these 4 stars.  Except for NaI, $\sigma \le  0.25$ dex for
each ion included, which is
comparable to the observational and modelling errors.  Because
only the NaD doublet lines were used, the Na abundance is highly
sensitive to errors in $T_{eff}$, and the large $\sigma$ may
be a reflection of this as well as of a potential range in abundance
of Na/Fe.

The abundances of Table~3 are in good agreement with those of
Cohen (1979) and of the most
recent published analyses by Peterson et al (1990),
Sneden et al (1991) and Armosky et al (1994).  
Sneden et al (1991) find [Fe/H] = $-2.25 \pm0.02$ dex from
9 red giants in M92, while we obtain $-$2.29 dex.  (The fifth and hottest
M92 blue HB star gives $-2.35$ dex, from FeII lines only.)
We therefore
are confident that our procedure is sound and that when we proceed
to analyze HB stars with similar $T_{eff}$ in much more metal-rich globular clusters the
results should be valid.

\begin{deluxetable}{lcrr}
\tablewidth{0pc}
\scriptsize
\tablecaption{Heavy Element Abundances
for Four Blue Horizontal
Branch Stars in M92}
\tablehead{
\colhead{Element} & \colhead{N$_{lines}$} & 
\colhead{[X]\tablenotemark{a}}  & 
\colhead{$\sigma$}\tablenotemark{b}  \nl
\colhead{  }  &  \colhead{  }  & \colhead{(dex)} &  \colhead{(dex)}  }
\startdata
NaI & 2 & $-$2.42 & 0.36 \nl
\noalign{\vskip 4pt} \nl
MgI & 4 & $-$1.92 & 0.22 \nl
MgII & 1{\tablenotemark{c}} & $-$1.95 & 0.07 \nl
\noalign{\vskip 4pt} \nl
CaI & 1 & $-$2.02 & 0.25 \nl
TiII & 9 & $-$1.90 & 0.18 \nl
CrII & 2 & $-$1.96 & 0.16 \nl
\noalign{\vskip 4pt} \nl
FeI & 5 & $-$2.34 & 0.19 \nl
FeII & 7 & $-$2.26 & 0.24 \nl
\noalign{\vskip 4pt} \nl
BaII & 1 & $-$2.50 & 0.25 \nl
\tablenotetext{a}{[X] = log10(X)(M92 star) -- log10(X)(Sun).}
\tablenotetext{b}{This is the standard deviation of the abundances
for the four stars about the mean, not the standard error of the mean
itself.}
\tablenotetext{c}{This is the close doublet of MgII at 4481\AA.}
\enddata
\end{deluxetable}

\begin{deluxetable}{lcr}
\tablewidth{0pc}
\scriptsize
\tablecaption{Possible Correlation of
Abundance and Rotation}
\tablehead{
\colhead{Star(s)} & \colhead{$v_{rot}(sin(i))$ Group} & 
\colhead{[$<\rm{A}>$]}   }
\startdata
X--22 & Low & $-0.26$  \nl
IV--27 + XII--9 & Intermediate & 0\tablenotemark{a}  \nl
XII--1 & High & +0.13  \nl
\tablenotetext{a}{Set to 0 by definition.}
\enddata
\end{deluxetable}

\subsection{Abundance and Rotation}

There is evidence for a small correlation between abundance
and projected rotational velocity.  (If this is true, then
the projected rotational velocity must be a good indicator of
the true rotational velocity and not just indicate inclination angle.)
As our best abundance indicator, we take the average of the
best determined elements, 
namely $\lbrack$[Mg](Mg I) + [Mg](Mg II)$\rbrack$,
[Ti](Ti II), and $\lbrack$[Fe](Fe I) + [Fe](Fe II)$\rbrack$.
The average of those three quantities we denote [$<\rm{A}>$].  This is
given in Table~4 for the star with the highest $v_{rot}(sin(i))$, the
two stars with intermediate rotation, and the star with the smallest
projected rotation.  The hottest M92 blue HB star is ignored.

The equivalent widths are of high accuracy here;
the accuracy of the $T_{eff}$ determinations is the limiting factor
in our ability to isolate small but real abundance
differences from star to star. 
The possible trend of increasing [A/H] with
increasing rotation urgently requires confirmation with a larger sample
of HB stars in M92. 

\section{SUMMARY}

We have analyzed high dispersion and high precision Keck/HIRES
spectra of 
5 blue horizontal
branch stars in the globular cluster M92 to 
establish that the projected rotational
velocity for these stars ranges from 15 to 40 \kms.  This is
larger than that expected based on the assumed rotation of their main
sequence progenitors, the spin down of rotation with age, and
the conservation of angular momentum.  Possible explanations
include a rapidly rotating stellar core.

An abundance analysis of these spectra of blue HB stars in M92 
yields the same results obtained from the giants in this cluster.
There is evidence of a small increase in abundance of the heavier
elements as $v_{rot}(sin(i))$ increases, but the sample urgently 
needs to be enlarged.

We now feel confident that our methods are valid and that we are
ready to proceed to analyze the spectra of HB stars in
much more metal-rich globular clusters.

\acknowledgements

We are grateful to R.Earle Luck for providing the computer-ready
compendium of {\it gf-}values as well as for the LTE model atmosphere
and abundance analysis codes provided by Bob Kurucz and Chris Sneden.
We are grateful to Mike Pahre for providing the infrared photometry
for the M92 blue HB stars.

It is a pleasure to thank the W.M.Keck Foundation, and its 
late President
Emeritus, Howard B. Keck, for the generous grant that made the Keck
Observatory possible.
We thank Theresa Chelminiak for observing assistance at the telescope.

\clearpage
\begin{figure}[ht]
\epsscale{1.0}
\plotone{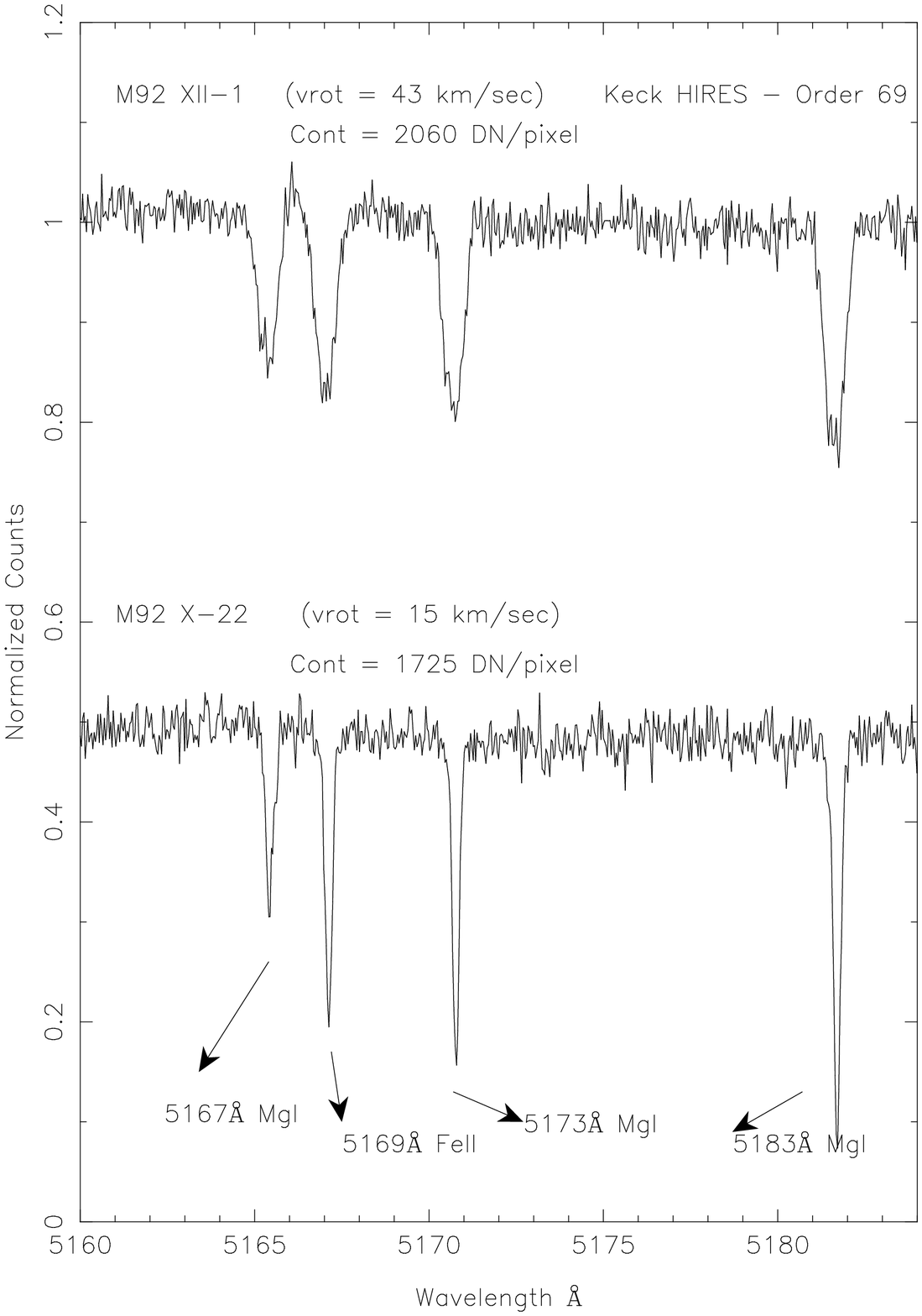}
\caption[Figure 1]{A 24\AA~section of spectrum from order 69 from
the HIRES spectrum of the blue HB stars XII--1 and X--22 in the
globular cluster M92.  The spectra have been normalized to the
continuum values indicated in each panel.  The spectrum of 
M92 X--22 has been vertically shifted by 0.5.  The projected
rotational velocities are indicated.
\label{fig1}}
\end{figure}
\clearpage

\begin{figure}[ht]
\epsscale{1.0}
\plotone{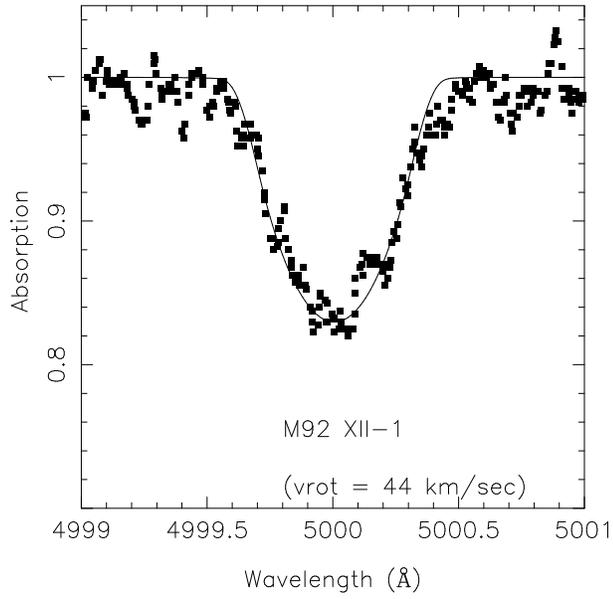}
\caption[Figure 2]{The points represent the
scaled, shifted and overlaid profiles of four absorption lines
in the blue HB star M92 XII--1, with the line center set to
5000 \AA~ in each case.  A four pixel running mean is shown.
The solid line is the profile of an Th-Ar
arc emission line broadened by a projected rotational velocity of 44 \kms.
\label{fig2}}
\end{figure}

\clearpage
\end{document}